\documentclass[preprint,12pt]{elsarticle}

\usepackage{graphicx}
\usepackage{dcolumn}
\usepackage{bm}
\usepackage[cp1250]{inputenc}
\usepackage{ifthen}
\usepackage{amssymb}

\journal{Physica C}

\begin{document}
\newcommand{\bk}{\mathbf{k}}
\newcommand{\bQ}{\mathbf{Q}}
\newcommand{\bq}{\mathbf{q}}
\newcommand{\bfr}{\mathbf{r}}
\newcommand{\ua}{\uparrow}
\newcommand{\da}{\downarrow}
\newcommand{\eq}{\begin{equation}}
\newcommand{\eqx}{\end{equation}}
\newcommand{\eqn}{\begin{eqnarray}}
\newcommand{\eqnx}{\end{eqnarray}}

\title{Andreev reflection between a normal metal and the FFLO superconductor II: a self-consistent approach}

\author[uj]{J Kaczmarczyk}
\ead{kaczek@gmail.com}

\author[uj]{M Sadzikowski}
\ead{sadzikowski@th.if.uj.edu.pl}

\author[uj,agh]{J Spałek}
\ead{ufspalek@if.uj.edu.pl}

\address[uj]{Instytut Fizyki im. Mariana Smoluchowskiego, Uniwersytet Jagielloński, Reymonta~4 30-059 Kraków, Poland}
\address[agh]{Wydział Fizyki i Informatyki Stosowanej, AGH, Reymonta 19, 30-059 Kraków, Poland}

\date{\today}

\begin{abstract}
We consider Andreev reflection in a two dimensional junction between a normal metal and a heavy fermion superconductor in the Fulde-Ferrell (FF) type of the Fulde-Ferrell-Larkin-Ovchinnikov (FFLO) state. We assume $s$-wave symmetry of the superconducting gap. The parameters of the superconductor: the gap magnitude, the chemical potential, and the Cooper pair center-of-mass momentum $\bQ$, are all determined self-consistently within a mean-field (BCS) scheme. The Cooper pair momentum $\bQ$ is chosen as perpendicular to the junction interface. We calculate the junction conductance for a series of barrier strengths. In the case of incoming electron with spin $\sigma = \ua$ only for magnetic fields close to the upper critical field $H_{c2}$, we obtain the so-called \textit{Andreev window} i.e. the energy interval in which the reflection probability is maximal, which in turn is indicated by a peak in the conductance.
The last result differs with other non-self-consistent calculations existing in the literature.
\end{abstract}

%\pacs{74.45.+c, 71.27.+a, 71.10.Ca, 74.50.+r}

%\pacs{
%74.45.+c  - Proximity effects; Andreev reflection; SN and SNS junctions\\
%71.27.+a  - Strongly correlated electron systems; heavy fermions \\
%71.10.Ca  - Electron gas, Fermi gas \\
%74.50.+r  - Tunneling phenomena; Josephson effects}


\maketitle

\section{Introduction}

Fulde-Ferrell-Larkin-Ovchinnikov (FFLO) superconducting state has been proposed theoretically in the early 1960s \cite{FF, LO}. In this unconventional superconducting state the Zeeman splitting for electrons at the Fermi surface makes it favorable for the Cooper pair to have a nonzero total momentum $\bQ = 2 \bq$ and consequently, the phase of the superconducting gap parameter oscillates spatially with the wave vector $\bQ$. By forming such condensate of moving Cooper pairs, the superconducting state survives to magnetic fields higher than the Pauli $H_{c2}$ limit. The FFLO state has suddenly gained renewed interest recently because of its possible realization in the heavy fermion superconductor CeCoIn$_5$ \cite{Koutroulakis, Kakuyanagi}, as well as in the layered organic superconductors $\kappa$-(BEDT-TTF)$_2$Cu(NCS)$_2$ \cite{Singleton} and $\beta''$-(ET)$_2$SF$_5$CH$_2$CF$_2$SO$_3$ \cite{Cho}. All those systems have a reduced dimensionality, what is crucial for FFLO phase stability, as then the orbital effects are suppressed and the Pauli effect (Zeeman splitting) may become the dominating factor. The FFLO state is also investigated in context of its possible realization in high density quark and nuclear matter \cite{Casalbuoni}, as well as in optical lattices~\cite{FFoptical}.

The irrefutable evidence for the FFLO state should be based on the phase sensitive experiments, because they can reveal the spatial variation of the phase or the sign of the order parameter. One of such experiments is the conductance spectroscopy in a normal metal (N) - superconductor (S) junction (NSJ). The method for analysis of the NSJ conductance has been provided by Blonder, Tinkham, and Klapwijk \cite{BTK}. Such experiments have already turned out to be successful for example in the determination of the gap parameter $d_{x^2 - y^2}$ symmetry in the CeCoIn$_5$ system \cite{Park}. A crucial role in the conductance spectrum is played by the Andreev reflection (AR) processes \cite{Andreev}. In the simplest view of the Andreev reflection an incident electron entering from N into S is converted at the NSJ interface into a hole moving in the opposite direction and Cooper pair inside SC.

The conductance characteristics for a NSJ with superconductor in the FFLO state has already been investigated for both the cases of FF (with $\Delta(\bfr) = \Delta_\bQ e^{i\bQ \bfr}$) \cite{Cui} and LO ($\Delta(\bfr) = \Delta_\bQ \cos(\bQ \bfr)$) \cite{Tanaka2} types of the FFLO state, as well as for the case of superconductor with supercurrent \cite{Zhang, Lukic} (i.e. the situation similar from formal point of view). The emphasis in these papers is put on the case with the $d$-wave symmetry of the gap (see also \cite{Tanaka1, Bruder, Kashiwaya, Argyropoulos} for the case of a BCS state with $d$-wave symmetry).

Here we consider an $s$-wave superconductor in the FFLO state of the FF type with the $\bq$ vector oriented perpendicular to the junction interface (similarly to Ref. \cite{Tanaka2} and different from Ref. \cite{Cui} in which $\bq$ is parallel to the interface). Recently, similar investigation has been performed for a one-dimensional situation \cite{Partyka} and it has been shown that with superconductor in the FF state there are lower and upper bounds on energy of the incoming electron for the Andreev reflection to take place (an Andreev window appears). Aim of the present paper is to analyze the Andreev window in a {\it self-consistent model}. Namely, we choose such Cooper pair momentum $2\bq$, which minimizes the free energy of the system (contrary to Refs. \cite{Tanaka2, Partyka}, in which $\bq$ is fixed) and we adjust the chemical potential $\mu$ so that the particle number $n$ is kept constant. We show that such careful examination of the superconductor properties leads to important alterations of the conductance spectrum.

We analyze the situation for heavy quasiparticles, because the FFLO state is more likely to appear when the carriers are heavy (then the orbital-effect influence is suppressed). For heavy fermion systems it is peculiar why AR takes place at all. Based on the BTK theory due to a large Fermi-velocity mismatch it should be severely limited by a high effective barrier strength $Z$. Nevertheless, AR is observed in those systems and theoretical efforts have been made to understand why it is the case \cite{Deutscher, Araujo, Araujo2}. Here we neglect the Fermi velocity mismatch by assuming equal masses of quasiparticles and equal chemical potentials on both sides of the junction.

The structure of the paper is as follows. In Section \ref{sec:sc} we discuss the superconducting state of quasiparticles with heavy effective masses in both two- (2D) and three-dimensional (3D) cases. In Section \ref{sec:AR} we present the results concerning Andreev reflection in a normal metal - heavy fermion superconductor junction. Finally, in Section \ref{sec:summary} we provide a summary and outlook.

\section{Fulde-Ferrell superconducting state characteristics} \label{sec:sc}

We consider here 2D and 3D system of paired quasiparticles with heavy masses $m~=~100\,m_e$, which roughly corresponds to the heaviest band of CeCoIn$_5$ \cite{McCollam}. We assume $s$-wave pairing symmetry, which has not been analyzed in detail as yet, since most of condensed matter correlated-electron systems hosting FFLO state are believed to have a $d$-wave gap symmetry.

The system of self-consistent equations describing the superconducting state is very similar to the one presented in \cite{JKJS} (see Section V) except that here we do not take into account two characteristic features for strongly-correlated electrons: the spin-dependence of quasiparticle mass and the effective field acting on them. These features appear in both slave-boson and Gutzwiller approaches to correlated electrons, but in the latter only if the Gutzwiller band-narrowing factors are taken as dependent on the system spin-polarization and supplemented with Maximum Entropy based treatment \cite{JJJK}. In this respect, we analyze here the more standard situation and the case with spin-dependent masses and effective field is left to a separate study.

Explicitly, the free energy functional, the gap parameter $\Delta_\bQ$ for the center-of-mass-momentum $\bQ$, and the condition for the number of particles $n$ (per atom) have the respective forms:
\eqn
 \mathcal{F} &=& - k_B T \sum_{\bk \sigma} \ln (1 + e^{-\beta E_{\bk \sigma}}) + \sum_\bk (\xi^{(s)}_\bk - E_\bk) + N \frac{\Delta^2}{V_0} + \mu N, \label{eq:sc1} \\
 \Delta_\bQ &=& \frac{V_0}{N} \sum_\bk \frac{1 - f(E_{\bk \uparrow}) - f(E_{\bk \downarrow})}{2 E_\bk} \Delta_\bQ, \label{eq:sc2} \\
 n &=& n_\uparrow + n_\downarrow = \frac{n}{N} \sum_{\bk \sigma} \Big\{u_\bk^2 f(E_{\bk \sigma}) + v_\bk^2 \big[1 - f(E_{\bk, -\sigma})\big]\Big\}, \label{eq:sc3}
\eqnx
where $\mathcal{F}(T, H_a; n, \Delta_\bq)$ is the system free-energy functional for the case of a fixed number of particles \cite{Koponen} (we set the band filling $n=0.97$), $V_0$ is the interaction potential, $u_\bk$, $v_\bk$ are the Bogolyubov coherence coefficients, $f(E_{\bk \sigma})$ is the Fermi distribution, and $n_\sigma$ is the spin-subband filling. The physical solution is that with a particular $\bQ$ which minimizes (\ref{eq:sc1}). The state with $\bQ = 0$ is called the BCS state, and that with $\bQ \neq 0$ - the FF state.

\begin{figure}[h!]
\includegraphics[angle=270, width=14cm]{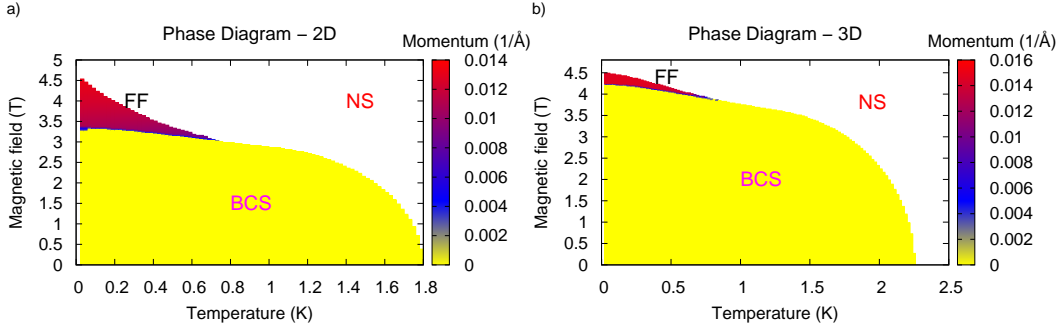}
\caption{\label{fig1} (Colour online). Applied magnetic field-temperature phase diagram for a 2D (a) and 3D (b) system of paired heavy quasiparticles. For further analysis of the Andreev reflection we take the parameters obtained along the $T=0.02 K \approx 0$ line. Note the extended regime of the FF state stability in the 2D case. This is only due to geometrical reasons, as we disregard orbital effects.}
\end{figure}

The quasiparticle spectrum is characterized by the energies (c.f. also \cite{Shimahara})
\eqn
E_{\bk \sigma} & \equiv & E_\bk + \sigma \xi^{(a)}_\bk, \quad \quad \quad \quad \quad \quad E_\bk \equiv \sqrt{\xi^{(s)2}_\bk+\Delta_\bQ^2}, \label{eq:Ek1} \\
 \xi^{(s)}_\bk & \equiv & \frac{1}{2} (\xi_{\bk + \bq \uparrow} + \xi_{-\bk + \bq \downarrow}), \quad  \xi^{(a)}_\bk \equiv \frac{1}{2} (\xi_{\bk + \bq \uparrow} - \xi_{-\bk + \bq \downarrow}), \label{eq:Ek}
\eqnx
with the dispersion relation taken in the computation as
\eq
\xi_{\bk\sigma} = \frac{\hbar^2 \bk^2}{2 m} - \sigma h - \mu, \label{eq:disp}
\eqx
%
%The pairing potential is equal to $V_0 = 110 \, K$ (2D) or $V_0 = 90 \, K$ (3D).
%
where $h\equiv g \mu_B H_a$ and $H_a$ is the applied magnetic field. Exemplary phase diagrams obtained for these parameters on the applied field $H_a$ and temperature $T$ plane are exhibited in Fig. \ref{fig1}.

\section{Andreev reflection: self-consistent solution} \label{sec:AR}

For the analysis of the Andreev reflection process we take the parameters obtained self-consistently (from the procedure presented above) for the superconducting state. Although this state is either 2D or 3D we consider only 2D NSJ for simplicity. Kinematics of the reflection may be analyzed by means of the Bogolyubov-de Gennes (BdG) equations
\eq
\left( \begin{array}{cc}
H_{0\hat{\sigma}}(\bfr) & \Delta(\bfr) \\
\Delta^*(\bfr) & -H_{0\hat{\sigma}}(\bfr)
\end{array} \right) \psi(\mathbf{r}, \sigma) =  E \psi(\mathbf{r}, \sigma),
\eqx
where the two-component wave function is given by
\eq
\psi(\bfr, \sigma) \equiv \left( \begin{array}{c} u(\bfr) | \sigma \rangle \\ v(\bfr) |\overline{\sigma}\rangle \end{array} \right),
\eqx
where $\sigma = \pm 1$ is the spin quantum number and $\overline{\sigma} = -\sigma$. The single-particle Hamiltonian is the following
\eq
H_{0\hat{\sigma}}(\bfr) = - \frac{\hbar^2 \nabla^2}{2 m} - \hat{\sigma} h - \mu + U(\bfr),
\eqx
where $\bfr = (x, y)$ and the interface scattering potential is chosen as a delta function of strength $H$, i.e. $U(\bfr)~=~H~\delta(x)$.

\begin{figure}[h!]
\scalebox{1.4}{\includegraphics[angle=0]{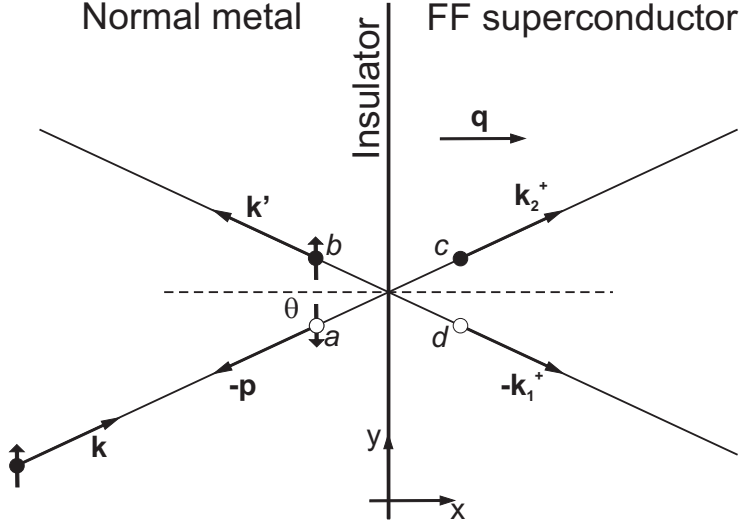}}
\caption{\label{fig2} Junction geometry for incoming particle of spin $\sigma = \ua$. Normal-state and Fulde-Ferrell regions are marked. Interface lies at the $x=0$ line. Full circles mark quasiparticles and empty ones mark quasiholes. Momentum of each of them is marked with a boldface letter, and amplitude with an italic letter. Namely, incoming particle has the momentum $\bk$, and amplitude 1, reflected hole has $\bf{p}$ and ${\it a}$, reflected quasiparticle: $\bk'$, ${\it b}$, transmitted quasiparticle: $\bf{k_2^+}$, ${\it c}$, and transmitted quasihole: $\bf{k_1^+}$, ${\it d}$. The angle of incidence is equal to $\theta$ and to the angle of reflection but other angles (of reflection of quasihole and those of transmissions) may differ by a small amount (c.f. also Fig. \ref{fig4}).}
\end{figure}

\begin{figure}
\includegraphics[angle=270,width=15cm]{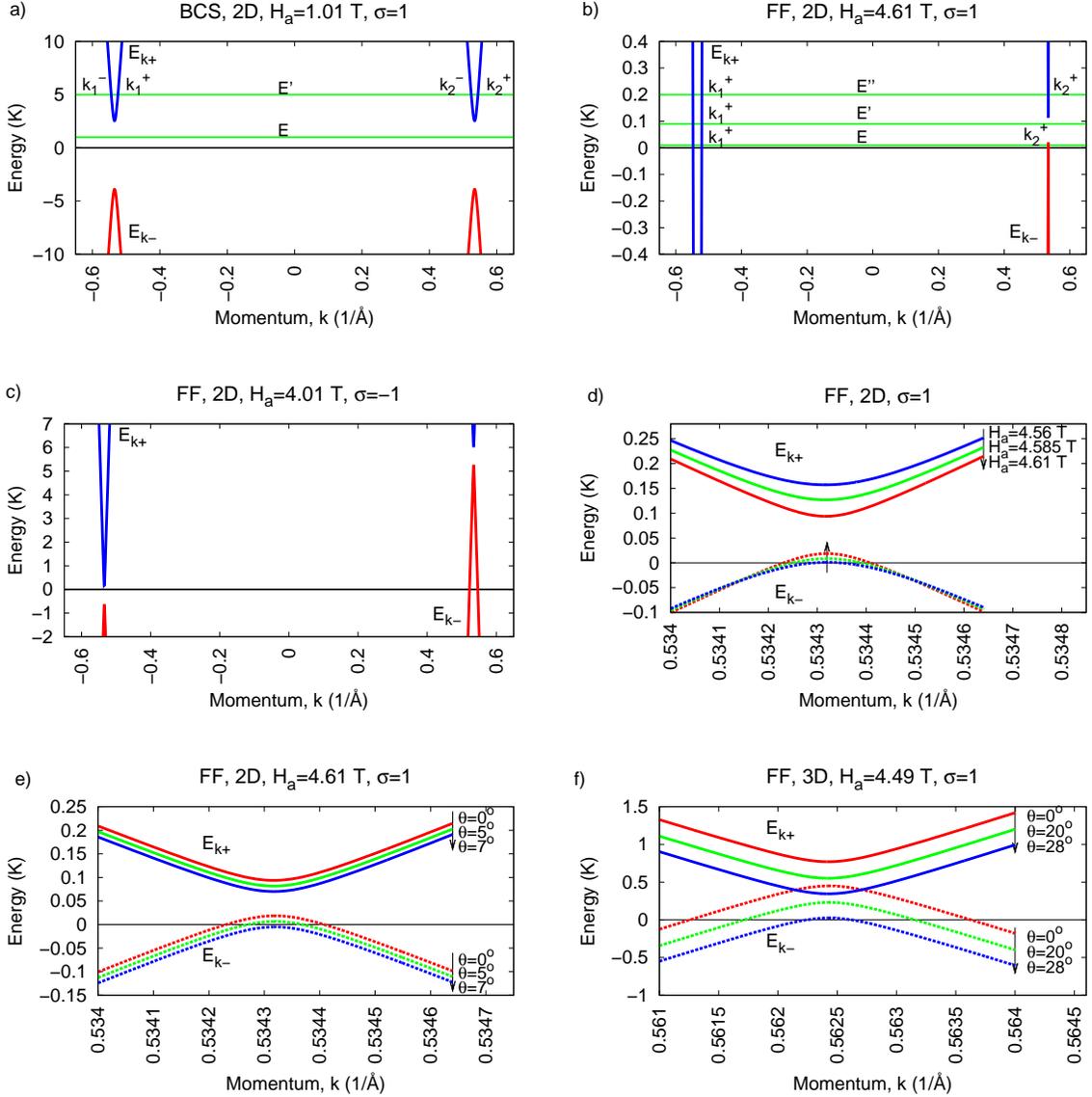}
\caption{\label{fig3} (Colour online). Quasiparticle and quasihole dispersion relations $E_{\bk \pm}$: (a) for BCS state at $H_a=1.01\textrm{ T}$; (b), (c) for FF state at $H_a=4.61\textrm{ T} \lesssim H_{c2}$ (b) and $H_a=4.01\textrm{ T}$ (c) for incoming electron with spin $\sigma=1$ (b) and $\sigma=-1$ (c); (d-f) closer view at the region around $k \approx + k_F$ with magnetic field close to $H_{c2}$. (d) shows $E_{\bk \pm}$ for a set of magnetic field $H_a$ values; (e) and (f) show $E_{\bk \pm}$ for a choice of angle of incidence $\theta$ values for 2D (e), and 3D (f) superconducting states. The angle of incidence is $\theta = 0$ in (a)-(d). The quasimomenta $k_{1,2}^\pm$ marked in (a) and (b) are solutions to the equations $E = E_{\bk\pm}$ propagating in the positive (superscript "$+$") and negative ("$-$") $x$ direction. Situation with incident particle with energy $E$ in (a) ($E'$ in (b)) results in Andreev reflection process. Note that the dispersion relations for BCS (a) do not depend on the incident angle $\theta$.
}
\end{figure}

We choose the Fulde-Ferrell (FF) type of superconducting state, in which $\Delta(\bfr)~=~\Delta_\bQ e^{i 2 \bq \bfr}$ and set the direction of the Cooper pair momentum $\bQ = 2\bq$ as perpendicular to the junction interface (similarly as in Refs. \cite{Tanaka2} and \cite{Partyka}), namely $\bQ = (Q, 0)$. The junction geometry is presented in Fig.~\ref{fig2}. The plane wave ansatz
\eq
\psi(\bfr, \sigma) = e^{i \bk \bfr} \left( \begin{array}{c} \tilde{u} e^{i \bq \bfr} \, | \sigma \rangle \\ \tilde{v} e^{-i \bq \bfr} \, |\overline{\sigma}\rangle \end{array} \right),
\eqx
with $\tilde{u}$ and $\tilde{v}$ as constants, leads to the following matrix equation
\eq
\left( \begin{array}{cc}
-E + \xi_{\bk+\bq, \sigma} & \Delta_\bQ \\
\Delta_\bQ^* & -E - \xi_{\bk-\bq, \overline{\sigma}}
\end{array} \right) \left( \begin{array}{c} \tilde{u} \, | \sigma \rangle \\ \tilde{v} \, |\overline{\sigma}\rangle \end{array} \right) =  0, \label{eq:matrix}
\eqx
where quasiparticle energies are given by (\ref{eq:disp}). Equation (\ref{eq:matrix}) gives the dispersion relations for quasiparticles and quasiholes in the superconductor
\eq
E_{\bk\pm} = \left\{ \begin{array}{c} \xi_\bk^{(a)} \pm \sqrt{\xi_\bk^{(s)2} + \Delta_\bQ^2}, \quad \textrm{for } \sigma = \ua \\
    \xi_{-\bk}^{(a)} \pm \sqrt{\xi_{-\bk}^{(s)2} + \Delta_\bQ^2}, \quad \textrm{for } \sigma = \da \end{array} \right.
\label{eq:Epm}
\eqx
where $\xi_\bk^{(s,a)}$ have been defined earlier in (\ref{eq:Ek}). One may check that the above equation is in accordance with (\ref{eq:Ek1}), as $E_{\bk+} = E_{\bk\ua}$ (quasiparticle) and $E_{\bk-} = - E_{\bk\da}$ (quasihole) for incoming particle with spin $\sigma = \ua$, as well as $E_{\bk+} = E_{-\bk\da}$ (quasiparticle) and $E_{\bk-} = - E_{-\bk\ua}$ (quasihole) for incoming particle with spin $\sigma = \da$. The plots of the energies (\ref{eq:Epm}) are exhibited in Fig. \ref{fig3}.

As we consider electron injected from the conductor side of the junction, the corresponding wave functions can be expressed as (we have omitted the spin part for clarity)
\eqn
\psi_{<}(\bfr) = \left( \begin{array}{c} 1 \\ 0 \end{array} \right) e^{i\bk\bfr} + a \left( \begin{array}{c} 0 \\ 1 \end{array} \right) e^{i \bf{p} \bfr} + b \left( \begin{array}{c} 1 \\ 0 \end{array} \right) e^{i\bk'\bfr}, \\
\psi_{>}(\bfr) = d \left( \begin{array}{c} u_1 e^{i q x} \\ v_1 e^{-i q x} \end{array} \right) e^{ i\bk_1^{+} \bfr} + c \left( \begin{array}{c} u_2 e^{i q x} \\ v_2 e^{-i q x} \end{array} \right) e^{ i\bk_2^{+} \bfr},
\eqnx
where $\psi_{<}(\bfr)$ and $\psi_{>}(\bfr)$ describe wave function on the normal-metal and superconductor sides, respectively. The quasimomenta $\bk_1^+$ (for quasihole) and $\bk_2^+$ (for quasiparticle) are solutions of (\ref{eq:Epm}) for a given incident energy $E$ propagating in the positive x direction. From the translational symmetry of the junction along the $y$ direction comes conservation of the $y$ momentum component. Namely, $k_y = k'_y = p_y = k_{1y}^+ = k_{2y}^+$. All the wave vectors are presented in Fig. 4.

\begin{figure}
\includegraphics[width=14cm]{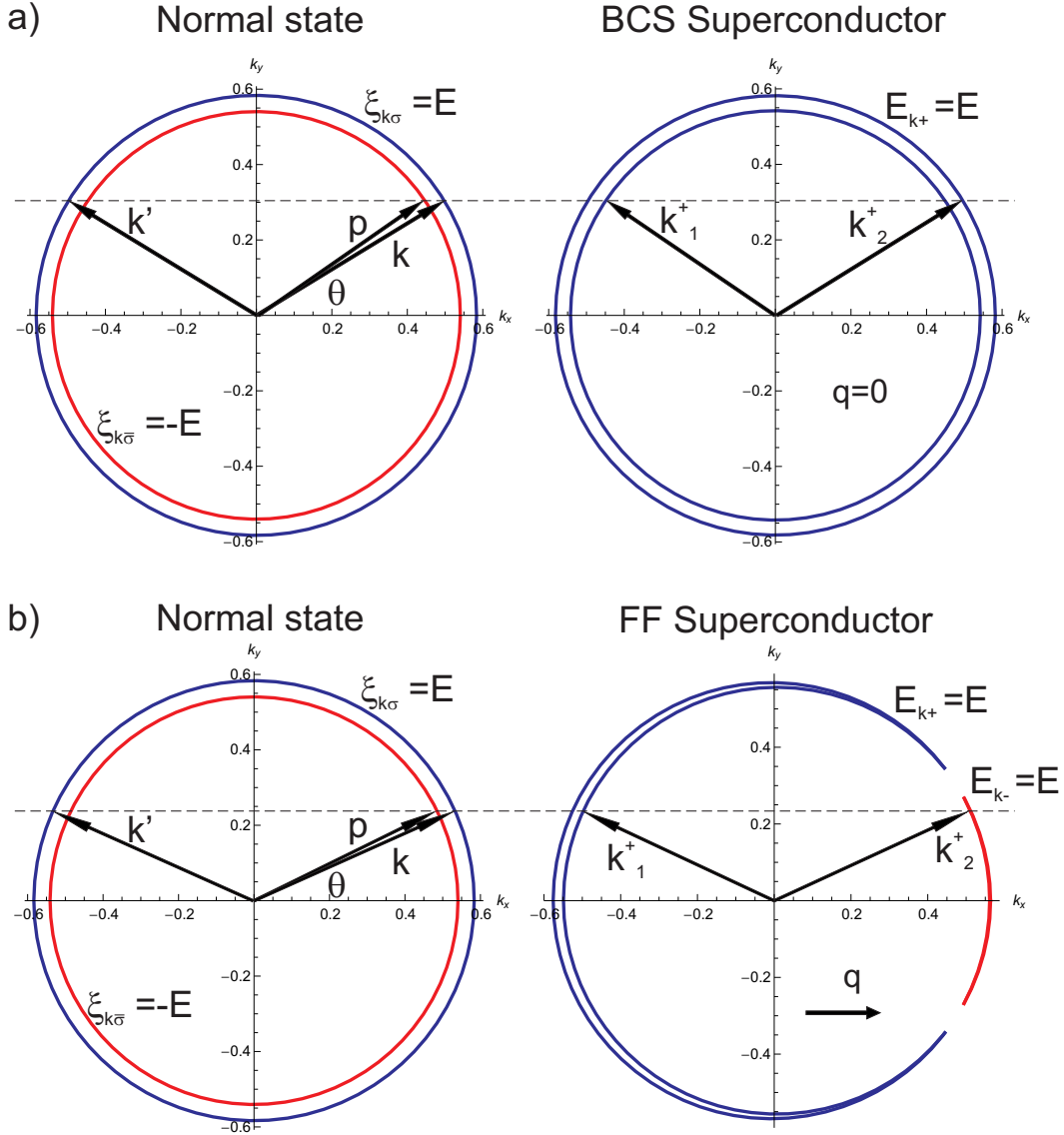}
\caption{\label{fig4} (Colour online). The junction geometry in the reciprocal space. All vectors are marked. It can be seen that only the incident and reflection angles are equal to $\theta$. It can be anticipated at this point that changing $\theta$ for BCS state does not lead to drastic changes in the transmission/reflection probabilities, whereas for the FF state the situation is quite different since $\bq \neq 0$ induces anisotropy in the reciprocal space. The energy $E$ value has been chosen as $10\,K$ for all graphs except (b) "FF Superconductor" for which $E=0.01\,K$ (for $E>0.5 \,K$ there would be no $E=E_{\bk-}$ regions in this case). The dashed lines are guide to eye and illustrate the conservation of momentum $y$-component.
}
\end{figure}
\begin{figure}
\includegraphics[angle=270, width=15cm]{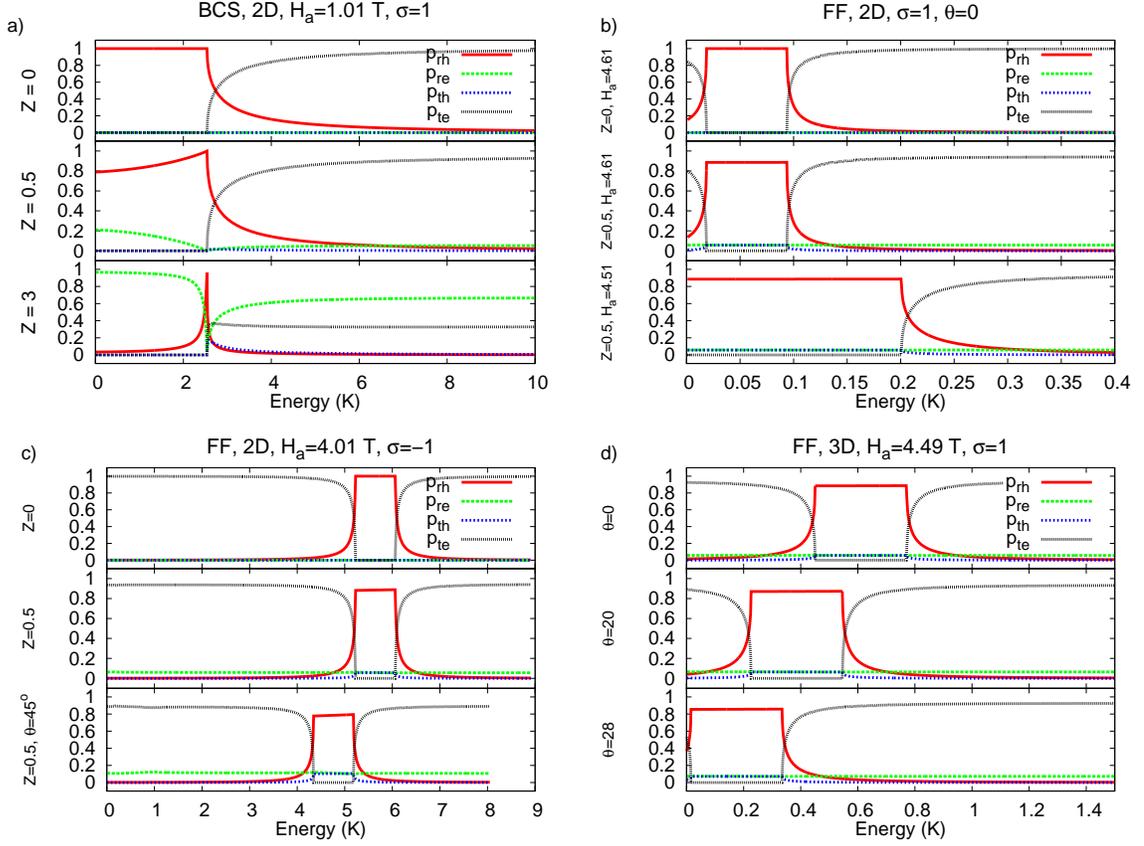}
\caption{\label{fig5} (Colour online). Probabilities of reflection and transmission processes (a) for BCS state and several barrier strength Z values; (b) for FF state for barrier strengths $Z=0$ (top), $Z=0.5$ (middle) and magnetic field $H_a = 4.51\textrm{ T}$ (bottom); (c) for FF state with incoming electron of spin $\sigma=-1$ and field $H_a=4.01\textrm{ T}$; (d) for 3D FF state for several angles of incidence $\theta$. The angle of incidence is taken $\theta = 0$ unless stated otherwise. Note that the probabilities for BCS state (a) do not depend on $\theta$.}
\end{figure}

We use standard boundary conditions
\eqn
\psi_<(\bfr)|_{x=0} = \psi_>(\bfr)|_{x=0}, \\
\frac{\partial \psi_<(\bfr)}{\partial x}|_{x=0} = \frac{\partial \psi_>(\bfr)}{\partial x}|_{x=0} - \frac{2 m H}{\hbar^2} \psi_<(\bfr)|_{x=0}.
\eqnx
Those conditions lead to the following set of 4 equations for the amplitudes ($a, b, c, d$)
\eqn
1 + b - c u_2 - d u_1 = 0, \\
a - c v_2 - d v_1 = 0, \\
i k_x (1 - b) - c u_2 i (q + k_{2x}^{+}) - d u_1 i (q + k_{1x}^{+}) + \frac{2 m H}{\hbar^2}(1 + b) = 0,\\
a i p_x - c v_2 i (k_{2x}^{+} - q) - d v_1 i (k_{1x}^{+} - q) + \frac{2 m H}{\hbar^2} a = 0,
\eqnx
which are very similar to those in e.g. \cite{Mortensen}, but for our case vectors are replaced by their $x$-components: e.g. $k \leftrightarrow k_x$, $p \leftrightarrow p_x$. From its solution one can obtain probabilities of hole reflection $p^\sigma_{rh} = |a|^2 \frac{\Re[p_x]}{k_x}$, particle reflection $p^\sigma_{re} = |b|^2$, quasiparticle transmission
\eq
p^\sigma_{te} = |c|^2 \frac{(|u_2|^2 - |v_2|^2) \Re[k^+_{2x}] + q}{k_x},
\eqx
and quasihole transmission
\eq
p^\sigma_{th} = |d|^2 \frac{(|u_1|^2 - |v_1|^2) \Re[k^+_{1x}] + q}{k_x},
\eqx
where the $\sigma$ superscript indicates the spin of the incoming electron. These probabilities are plotted in Fig. \ref{fig5} for various barrier strengths $Z~\equiv~2 m H/(k_F \hbar^2)$, where we define Fermi wave vector $k_F$ using the zero-field value $k_F = \frac{1}{\hbar} \sqrt{2 m \mu}$. Note also that we do not use the assumption $k = k' = p = k_1^+ = k_2^+ \approx k_F$ utilized at this point in majority of papers on Andreev reflection, because we deal with heavy quasiparticles for which $\mu$ is of the order of $100 \, K$. Therefore the assumption $\mu \gg E$ is not, strictly speaking, applicable in the present situation.

As can be seen in Fig. \ref{fig5}, there are lower and upper bounds on the energy of incident electron for the Andreev reflection to take place. For example in (b) this region is $[0.02 \, K, \, 0.1 \, K]$ and corresponds to the region in Figs. \ref{fig3}b and \ref{fig3}d between the maximum of $E_{\bk-}$ and minimum of $E_{\bk+}$. The position and existence of the Andreev window is strongly dependent on the parameters chosen. Namely, a deviation from $H_{c2}$ by $1\%$ (for a 2D system) or from the perpendicular incidence by 5 (2D) or 28 (3D) degrees makes it disappear (the lower bound on the energy E at which the Andreev reflection takes place goes below $E=0$). It is clear that the Andreev window persists to larger degrees for a 3D system (for geometrical and energetical reasons, c.f. also Fig. \ref{fig3}f). For incident electron with spin $\sigma=-1$ the situation is different as Andreev window appears for high $E$ values (e.g. between $5.2\, K$ and $6.1\,K$ - see Fig. \ref{fig3}c and Fig. \ref{fig5}c). Note also the qualitatively different behavior of the probabilities for the BCS (a) and FF (b-d) superconductor. Namely, the reflection of a hole probability $p_{rh}$ for FF is constant in the whole Andreev window region and decreases less rapidly with the increasing $Z$ (compare the middle graphs of Fig. \ref{fig5} (a) and (b)). This behavior produces very different conductance peaks for the two cases.

\begin{figure}
\includegraphics[width=15cm]{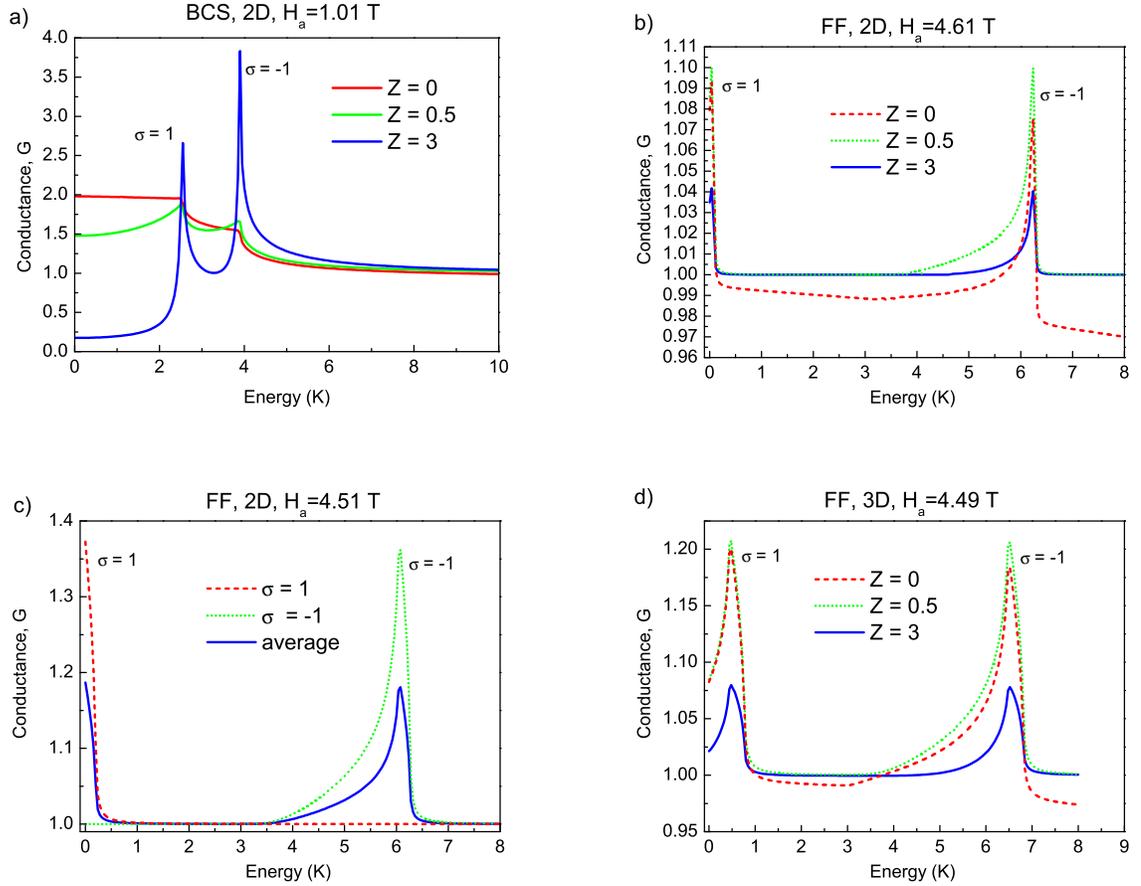}
\caption{\label{fig6} (Colour online). Conductance plots for different $Z$ values for (a) BCS state, (b) FF state at $H_a = 4.61\textrm{ T} \approx H_{c2}$, (c) FF state at a slightly lower field $H_a=4.51\textrm{ T}$ with the signals from both spin channels, and (d) for a 3D system at $H_a=4.49\textrm{ T} \approx H_{c2}$. Note that the distance between the peaks is twice the Zeeman energy $2h = 2 g \mu_B H_a$.
}
\end{figure}

Differential conductance ($G \equiv dI/dV$) can be obtained from the reflection and transmission probabilities \cite{BTK, Chaudhuri} in a straightforward manner
\eq
G_{ns}^\sigma = \frac{1}{2} \int_{-\pi/2}^{\pi/2} d\theta \cos{\theta} [1-p^\sigma_{re}(E, \theta) + p^\sigma_{rh}(E, \theta)]. \label{eq:G}
\eqx
The final result of our calculation is the conductance $G$ averaged over spin and normalized with respect to the conductance $G_{nn}^\sigma$ of the junction with $\Delta=0$. Namely,
\eq
G = \frac{G_{ns}^\ua + G_{ns}^\da}{G_{nn}^\ua+G_{nn}^\da}.
\eqx
This quantity is exhibited in Fig. \ref{fig6}. In Fig. \ref{fig6}c also the spin-resolved conductance $G^\sigma~\equiv~G_{ns}^{\sigma} / G_{nn}^{\sigma}$ is presented.

The Andreev window for $\sigma=\uparrow$ manifests itself by a peak at $E=0.02 K$ (b) and $E=0.5 K$ (d). This peak is not present in 2D case for $H_a \leq 4.535\textrm{ T}$ (see (c)) because then the quasihole energy $E_{\bk-}$ falls below $E=0$ (see Fig. \ref{fig3}) and there is no Andreev window for $\sigma=\uparrow$ in this case. Note that in the non-self-consistent calculations \cite{Tanaka2, Partyka} the $\sigma =\ua$ peak was fully present (c.f. Fig. 4a of Ref. \cite{Tanaka2}). Note also that (a) agrees qualitatively with the results of e.g. Ref \cite{Cui}. As should be expected the $\sigma = \ua$ peak is more pronounced for 3D system, but for 2D it is also present for the applied field close to $H_{c2}$.

From the comparison of (a) with (b-d) it is evident that for FF state the conductance peaks are much broader than in the BCS case. This is because of the anisotropy in the reciprocal space induced by $\bq \neq 0$ (see Fig. \ref{fig4}), which gives $\theta$-dependent position of Andreev window. Therefore, integration of the $[1-p^\sigma_{re}(E, \theta) + p^\sigma_{rh}(E, \theta)]$ factor over $\theta$ in (\ref{eq:G}) gives broader peaks in the FF case.

\section{Conclusions} \label{sec:summary}
In this paper we have provided a detailed analysis of the Andreev reflection from an FFLO superconductor (of the FF type) by analyzing it within a fully self-consistent scheme for the $s$-wave symmetry of the superconducting gap and heavy quasiparticles. The conditions for the appearance of the Andreev window are determined explicitly. The Andreev window leads to peaks in the conductance spectrum. The $\sigma=\ua$ peak should be clearly visible only near the upper critical field. This result can be contrasted with other non-self-consistent calculations existing in the literature \cite{Tanaka2, Partyka}. The calculated differential conductance as a function of external bias allows for an experimental verification of the results provided the FF state with an $s$-wave symmetry of the gap is detected.

\section*{Acknowledgements}
The work was supported by Ministry of Higher Education and Science, Grants Nos. N N202 128736 and N N202 173735. JK was also supported by the "Doctus" grant. Technical help from T. Partyka is acknowledged.

\end{document}